\documentclass[aps,prd,preprint,superscriptaddress,showpacs,tightenlines,nofootinbib]{revtex4}

\usepackage{graphicx}
\usepackage{dcolumn}
\usepackage{bm}

\newcommand{\vub}{\mbox{$|V_{ub}|$}}

\newcommand{\dtopiorklnu}{\mbox{$D^0 \to \pi^-(K^-) \ell^+ \nu$}} 
\newcommand{\dtopiorkmunu}{\mbox{$D^0 \to \pi^-(K^-) \mu^+ \nu$}} 
\newcommand{\dtopilnu}{\mbox{$D^0 \to \pi^- \ell^+ \nu$}} 
\newcommand{\btopilnu}{\mbox{$B^0 \to \pi^- \ell^+ \nu$}} 
 
\newcommand{\dtopienu}{\mbox{$D^0 \to \pi^- e^+ \nu$}} 
\newcommand{\dtopimunu}{\mbox{$D^0 \to \pi^- \mu^+ \nu$}} 
\newcommand{\dtoklnu}{\mbox{$D^0 \to K^- \ell^+ \nu$}}
\newcommand{\dtokenu}{\mbox{$D^0 \to K^- e^+ \nu$}}
\newcommand{\dtokmunu}{\mbox{$D^0 \to K^- \mu^+ \nu$}}
\newcommand{\dtokslnu}{\mbox{$D^0 \to K^{*-} \ell^+ \nu$}} 
\newcommand{\dtokpilnu}{\mbox{$D^0 \to (K\pi)^-  \ell^+ \nu$}} 
\newcommand{\dtorholnu}{\mbox{$D^0 \to \rho^- \ell^+ \nu$}} 

\newcommand{\dtoxlnu}{\mbox{$D^0 \to X \ell^+ \nu$}}

\newcommand{\mhlnu}{\mbox{$m_{hl\nu}$}} 
\newcommand{\phlnu}{\mbox{$|p_{hl\nu}|$}} 
\newcommand{\mhl}{\mbox{$m_{hl}$}} 
\newcommand{\phl}{\mbox{$\vec{p}_{hl}$}} 
\newcommand{\phad}{\mbox{${p}_{h}$}} 
\newcommand{\delm}{\mbox{$\Delta m$}}

\newcommand{\pmiss}{\mbox{${\vec{p}}_{\rm miss} $}}

\newcommand{\qs}{\mbox{$q^2$}}
\newcommand{\Rz}{\mbox{$R_0$}}
\newcommand{\mpole}{\mbox{$m_{\rm pole}$}}
\newcommand{\rfo}{ \mbox{${|f^{\pi}_{+}(0)| / |f^K_{+}(0)|}$}}

\newcommand{\gev}{\mbox{{\rm GeV}}}
\newcommand{\mev}{\mbox{{\rm MeV}}}
\newcommand{\bbar}{\mbox{$B\bar{B}$}}

\begin{document}

\preprint{CLNS 04/1876}       
\preprint{CLEO 04-06}         

\title{Study of the  Semileptonic Charm Decays {\boldmath$D^0 \to \pi^- \ell^+ \nu$} and {\boldmath$D^0 \to K^- \ell^+ \nu$} }

\author{G.~S.~Huang}
\author{D.~H.~Miller}
\author{V.~Pavlunin}
\author{B.~Sanghi}
\author{E.~I.~Shibata}
\author{I.~P.~J.~Shipsey}
\affiliation{Purdue University, West Lafayette, Indiana 47907}
\author{G.~S.~Adams}
\author{M.~Chasse}
\author{J.~P.~Cummings}
\author{I.~Danko}
\author{J.~Napolitano}
\affiliation{Rensselaer Polytechnic Institute, Troy, New York 12180}
% ...
\collaboration{CLEO Collaboration} 
\noaffiliation


\date{July 19, 2004}

\begin{abstract} 
We investigate the decays $D^0 \to \pi^- \ell^+ \nu$ and $D^0 \to K^- \ell^+ \nu$, where $\ell$ is $e$ or $\mu$, using approximately 7 ${\rm fb}^{-1}$ of data collected with the CLEO III detector.  We find $\Rz \equiv {\cal B}(\dtopienu)/{\cal B}(\dtokenu) = 0.082 \pm 0.006 \pm 0.005$.  Fits to the kinematic distributions of the data provide parameters describing the form factor of each mode.  Combining the form factor results and \Rz\ gives $|f^{\pi}_{+}(0)|^2 |V_{cd}|^2/|f^K_{+}(0)|^2 |V_{cs}|^2 =   0.038^{+0.006+0.005}_{-0.007-0.003}$.
\end{abstract}

\maketitle



The quark mixing parameters are fundamental constants of the weak interaction.  Measuring them also tests the unitarity of the quark mixing (CKM) matrix, which is sensitive to as yet undiscovered particles and interactions.  Semileptonic decays have provided most quark coupling data.  For these decays, the strong interaction binding effects, parameterized by form factors, are simplest to calculate; nonetheless, even here, form factor uncertainties can dominate the experimental uncertainties~\cite{Motiv}.  

We present a study of the decays \dtopilnu\ and \dtoklnu, where $\ell = e\ {\rm or}\ \mu$.  Charge conjugate modes are implied throughout this paper.  We measure the ratio of their branching fractions, $\Rz \equiv {\cal B}(\dtopienu)/{\cal B}(\dtokenu)$, and, for the first time for \dtopilnu, parameters describing their form factors.  The study of the \dtopilnu\ form factor is particularly interesting because it tests predictions for that of the closely related decay \btopilnu, which provides \vub. 

In the limit $(m_\ell/m_c)^2 = 0$, where $m_\ell$ and $m_c$ are the lepton and charm quark masses, the differential partial widths for \dtopilnu\ and \dtoklnu, in terms of the form factor $f_+(q^2)$, are
$$
\frac{d\Gamma}{d\qs}(D\to h \ell\nu) = \frac{G_F^2}{24\pi^3}p_{h}^3|V_{cd(s)}|^2  |f_+^{h}(q^2)|^2  .
$$
Here $h$ is $\pi$ or $K$, and \qs\ is the invariant mass squared of the lepton-neutrino system and ranges from $m_\ell^2$ to 2.98(1.88) \gev$^2$ for \dtopiorklnu.  To reduce the form factor sensitivity of \Rz\ and determine the \qs\ distributions, the yields are extracted in bins of \qs.


We use $e^+ e^- \to c \bar{c}$ events collected at and just below the $\Upsilon(4S)$ resonance with the CLEO III detector~\cite{CLEOIII}.  We use only runs with good lepton identification, which leads to slightly different, but overlapping, datasets for the electron and muon modes with integrated luminosities of 6.7 and 8.0 fb$^{-1}$ respectively.  

A major challenge for this analysis is the contamination of the \dtopilnu\ sample by \dtoklnu\ decays, which are about a factor of 10 more common.  The use of a Ring-Imaging Cherenkov detector (RICH) and specific ionization in the drift chamber ($dE/dx$) reduces this contamination dramatically by distinguishing $K$ from $\pi$ mesons.  The resulting efficiency and misidentification probability suppress misidentified \dtoklnu\ decays to 15\% of the \dtopilnu\ signal.

The analysis also benefits from the hermeticity of the detector, which enables us to substitute the missing momentum vector of each event for the neutrino momentum.  Within the active region, which covers 93\% of the solid angle,  we accept photons with energies above 50 \mev\ and detect over 92\% of charged particles with momentum above 75 \mev.  

$D^0$ candidates are reconstructed from lepton, hadron ($\pi$ or $K$), and neutrino combinations.  Electron candidates have $p_e>0.6$ \gev, lie within the barrel of the detector ($|\cos\theta\ |<0.8$, where $\theta$ is the angle between the track and the beam)
and have the expected calorimeter, RICH, and $dE/dx$ signals.
Muon candidates have $p_\mu>1.5$ \gev, lie within $0.1<|\cos\theta\ |<0.6$, penetrate at least 5 interaction lengths of material, and have the expected energy deposit in the calorimeter.
The hadron ($h$) must have electric charge opposite that of the lepton and satisfy strong $K$ or $\pi$ identification requirements.
The missing momentum of the event (\pmiss) provides the first estimate of the neutrino momentum: it is the negative of the net momentum of all charged particles and calorimeter showers (treated as photons) that are not associated with a track.  We require $\mhlnu>1.6$ \gev.

To improve the neutrino momentum resolution, we impose the mass constraint $\mhlnu=m_{D^0}$ (we use PDG ~\cite{PDG} masses throughout).  An ellipsoid of neutrino momenta satisfies this requirement;  we take the momentum that lies in the plane defined by \pmiss\ and \phl\  (hadron-lepton momentum), and has the smallest vector difference from the missing momentum.  This procedure reduces the full width at half-maximum of the neutrino momentum resolution from 0.8 to 0.45 GeV. 

The kaons and pions from properly reconstructed decays tend to have higher momentum than those in backgrounds, so we demand $\phad>0.5$ \gev.  We further require  $0.6<\mhl<1.85$ \gev\ and $\phlnu>2.25$ \gev\ (low and middle \qs\ bins) and $\phlnu>3.0$ \gev\ (high \qs\ bin).   Semileptonic $B$ decays and most Bhabha, 2-photon, and $\tau^+\tau^-$ backgrounds are suppressed by
imposing $0.20<R_2<0.85$, where $R_2$ is the ratio of Fox-Wolfram moments~\cite{R2}. Bhabha and 2-photon events are also suppressed by demanding $|\cos\theta_{\rm thrust}|<0.8$ for events in which a candidate $e^+$ ($e^-$) lies in the hemisphere opposite the incident $e^+$ ($e^-$) beam.  Here $\theta_{\rm thrust}$ is the angle between the thrust axis of the event and the beam.

We require that all $D^0$ candidates come from the decay $D^{*+} \to D^0 \pi^+$.
We reconstruct the $D^{*+}$ by pairing a pion with the appropriate charge (the ``soft'' pion, $\pi_{\rm s}$) with the $D^0$ candidate and then computing the mass difference between the $D^{*+}$ and the $D^0$ candidates, $\Delta m = m_{h\ell\nu\pi_{\rm s}} - \mhlnu$.  
The signal peaks in the region $\Delta m < 0.16$ \gev\ (the ``signal region'') with a root-mean-square width of about 10 \mev.  We use the $\Delta m$ distribution to extract the yields. 


About half the background in the signal region is composed of candidates in which the $\pi_{\rm s}^+$ comes from a $D^{*+}$ decay.  This background is troublesome because it peaks in $\Delta m$, albeit more broadly than the signal, and, for \dtopilnu, is comparable in size to the signal.  A Monte Carlo simulation~\cite{geant} shows that most of this ``peaking'' background in the \dtopilnu\ sample comes from \dtoklnu\ decays in which the $K$ is mistaken for a $\pi$ (8\%), or from candidates in which a lepton from \dtoklnu\ (44\%), \dtokslnu\ (32\%), \dtorholnu\ (9\%), or nonresonant \dtokpilnu\ (2\%) is paired with a random pion or one from the same decay.  The remaining half of the background does not peak because the $\pi_{\rm s}$ is not from a $D^*$ decay (the ``false-$\pi_{\rm s}$'' background).  For the more common \dtoklnu\ mode, the ratios of both the peaking and false-$\pi_{\rm s}$ background to signal are smaller by a factor of ten.  The peaking background comes primarily from \dtokslnu\ (66\%), \dtokpilnu\ nonresonant (6\%), and \dtopilnu\ (4\%).  


We divide the data into three \qs\ bins: [0, 0.75] (bin 1), [0.75, 1.5] (bin 2), and $>1.5$ \gev$^2$ (bin 3).  The bin size is guided by our \qs\ resolution of 0.4 \gev$^2$.   To calculate \qs\ for \dtoklnu, we use $m_\pi$ in place of $m_K$ so that the \dtoklnu\ yield in each bin corresponds to the \dtoklnu\ background in the same \dtopilnu\ bin.  

\begin{figure}
\begin{center}
\includegraphics*[width=3.35in]{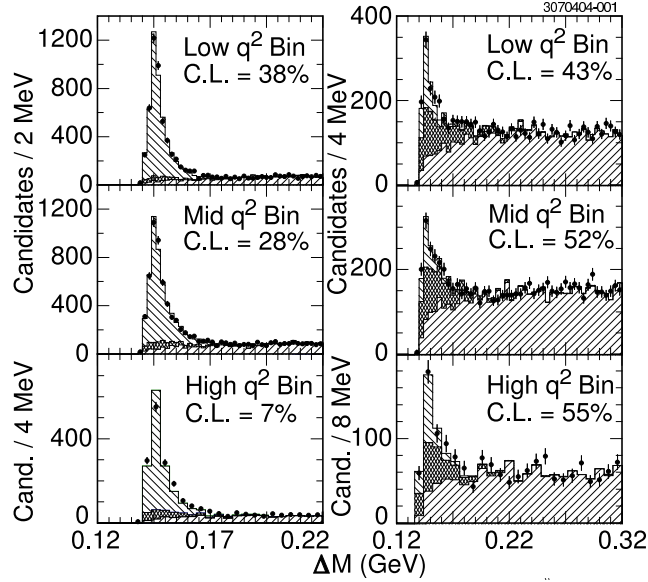} 
\caption{The fits to the $\Delta m$ distributions for \dtokenu\ (left) and \dtopienu\ (right) and their confidence levels (C.L.).  The data (points) are superimposed on the sum of the normalized simulated signal (peaked histogram), peaking background (dark histogram) and false-$\pi_{\rm s}$ background (broad histogram).}
\label{fig:fits}
\end{center}
\end{figure}

The yield in each \qs\ bin for each of the modes, \dtokenu, \dtokmunu, \dtopienu, and \dtopimunu, is determined from a fit to the $\Delta m$ distribution.  The Monte Carlo simulation~\cite{geant} provides the $\Delta m$ distributions of the signal and backgrounds.  The \dtoklnu\ samples are fit first.  The two free parameters in these fits are the normalizations of the \dtoklnu\ simulated signal and of the false-$\pi_{\rm s}$ background relative to the data.  Since the fit can only weakly distinguish the signal from the peaking backgrounds, we fix their ratio to the value predicted by the Monte Carlo simulation.  (This assumption is investigated in the section on systematic uncertainties.)   Then the  \dtopilnu\ samples are fit.   The normalization of \dtoklnu\ from the \dtoklnu\ fits sets the normalization of the peaking background in the \dtopilnu\ fits.  The two free parameters in these fits are the normalizations of the \dtopilnu\ signal and of the false-$\pi_{\rm s}$ background.   The electron mode fits and their confidence levels are shown in Figure~\ref{fig:fits}.  The muon fits are similar, but with smaller sample sizes because of the muon momentum and angular restrictions.  To test for sensitivity to the details of the fitting shape, we reanalyze the \dtokenu\ sample using \delm\ calculated with the uncorrected neutrino momentum and also using \delm\ calculated omitting the neutrino.  We see no significant variation in the results.

An efficiency matrix relates the number of decays produced in each \qs\ bin (the efficiency corrected yields) to the number detected in each bin.  Calculated using a Monte Carlo simulation, it accounts for both reconstruction efficiency and event migration across bins.  The average reconstruction efficiency for \dtopienu, not including the $D^{*+}\to D^0\pi_{\rm s}^+$ branching fraction, is about 11\%, and 30\% of reconstructed events migrate from their true \qs\ bin into another bin.   For \dtokenu, the migration is somewhat greater because we use the pion mass to compute \qs. 
Efficiencies are lower by a factor of four (six) for \dtopiorkmunu.

\begin{figure}
\begin{center}
\includegraphics*[width=3.4in]{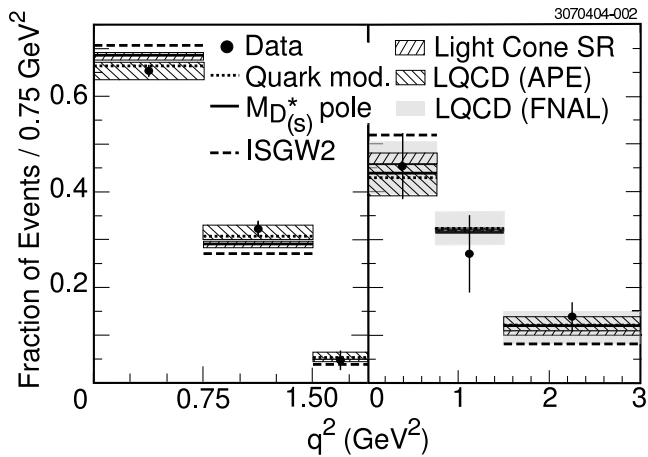} 
\caption{Distributions in \qs\ for \dtoklnu\ (left) and \dtopilnu\ (right), after correcting for reconstruction efficiency and smearing in \qs.  The data include statistical and systematic uncertainties and are overlaid with various predictions~[6-10]. }
\label{fig:qsqrDist}
\end{center}
\end{figure}

We sum the efficiency corrected yields over \qs\ bins to find $R_{0_e} = 0.085 \pm 0.006 \pm 0.006$ and $R_{0_\mu}= 0.074 \pm 0.012 \pm 0.006$ for the electron and muon modes, respectively, where the first uncertainty is statistical, and the second is systematic and is described below.  We then compute the normalized \qs\ distributions, which are defined as the fraction of the total corrected yield in each \qs\ bin (since the $D^*$ production rate is undetermined).  They are shown in Figure~\ref{fig:qsqrDist} and Table~\ref{tab:normy}.  The results combine the electron and muon modes after correcting the muon modes for their reduced phase space.   The correlations between the \qs\ bins are $\rho_{12}^K = -0.81$, $\rho_{13}^K = 0.18$, $\rho_{23}^K = -0.72$ and $\rho_{12}^\pi = -0.67$, $\rho_{13}^\pi = -0.23$, $\rho_{23}^\pi = -0.57$.

\begin{table}
\centering
\caption{The \qs\  bin yields, with statistical and systematic uncertainties, after correcting for efficiency and smearing across bins and normalizing their sum to unity. }  
\begin{tabular}{ccc}\hline\hline
 \qs\ ($\gev^2$) & $\Delta\Gamma/\Gamma (K^-\ell^+\nu)$ &$\Delta\Gamma/\Gamma  (\pi^-\ell^+\nu)$ \\\hline
$[0, 0.75]$   & $0.654\pm 0.010\pm 0.005$  & $0.45\pm0.05\pm0.03$  \\
$[0.75,1.5]$  & $0.323\pm0.015\pm0.006$    & $0.26\pm0.06\pm0.04$ \\
$>1.5$   & $0.024\pm0.008\pm0.006$    & $0.29\pm 0.05\pm0.02$  \\\hline\hline
\end{tabular}
\label{tab:normy}
\end{table}

The systematic uncertainties, summarized in Table~\ref{tab:syst}, are dominated by uncertainties in the backgrounds.  Inaccuracies in the simulation can affect the reconstructed neutrino momentum, thereby shifting the expected amount of peaking background relative to the \dtoklnu\ yield, and hence the extracted \dtopilnu\ yield.  To study such effects, we adjust variables in the simulation ($K_L$ production, tracking efficiency, track parameters, and shower energy resolution).  The sizes of these variations are guided by independent studies of the detector and the scale of the small discrepancies observed between data and simulated distributions in $\mhl$, $\mhlnu$ and $p_\nu$.  Biases in the simulation can also affect the \dtopilnu\ and \dtoklnu\ efficiency ratio and \qs\ distributions.  In practice these effects are small since the same selection criteria are applied to both modes and remaining differences depend primarily on the decay kinematics, which are readily simulated.  We find a small contribution from the uncertainties in the efficiencies for successfully identifying hadrons ($\pi$ or $K$) and leptons. 

Hadron misidentification, particularly mistaking a kaon from \dtoklnu\ for a pion from \dtopilnu, poses a serious problem.  The probability of misidentifying a kaon as a pion is measured as a function of momentum with a sample of $D^0\to K^-\pi^+$ decays.  The momentum-averaged misidentification probability is $(1.9\pm 0.1 {\rm (stat.)})$\%.  We test for differences in the misidentification probabilities between kaons from $D^0\to K^-\pi^+$ (where a tight mass cut is applied) and kaons in our sample (where the mass cut is very loose) by applying our technique for measuring misidentification probabilities to simulated events of both kinds, and see no hint of bias.  However, we see run-to-run variations in the misidentification probability that approach statistical significance, and accordingly assign it a conservative 20\% relative systematic uncertainty.

Additional uncertainty arises from the statistical uncertainty in the \dtoklnu\ normalization, since it determines the background level for \dtopilnu.  The branching ratios of other semileptonic modes, \dtoxlnu, relative to \dtoklnu\ also affect the yields, as do the form factors  for $D\to K^*\ell\nu$, the charm fragmentation parameters, the background from candidates in which a hadron is mistaken for an electron or muon, and the normalization of residual \bbar\ events.  

\begin{table}[ht]
\centering
\caption{The percent uncertainties in \Rz\ and the normalized raw \qs\ bin yields.  Entries are explained in the text.  Systematic uncertainties apart from the \dtoklnu\ normalization and a portion of the simulation uncertainty (first row) are correlated between the $\pi$ and $K$ modes. }
\begin{tabular}{l r r r r r r r}\hline\hline
Unc. Source  &  $\sigma_{R_0}$ & $\sigma_{\rm 1}^{\pi}$ &
$\sigma_{\rm 2}^{\pi}$ & $\sigma_{\rm 3}^{\pi}$	& $\sigma_{\rm 1}^{K}$ & $\sigma_{\rm 2}^{K}$ & $\sigma_{\rm 3}^{K}$ \\\hline
Simulation$^\dag$         &  2.9  & 3.4 & 4.1 & 6.0 & 0.5 & 0.5 & 2.6 \\
Part. ID ($e$)$^\dag$       &  1.9  & 0.7 & 0.8 & 1.3 & 0.3 & 0.3 & 0.4 \\
Part. ID ($\mu$)$^\dag$   &  2.0  & 0.9 & 0.9 & 1.0 & 0.4 & 0.4 & 0.6 \\
$K$/$\pi$ mis-ID$^\ddag$     &  3.9  & 3.1 & 3.5 & 3.2 & 0.0   & 0.0   & 0.0   \\
$K e\nu$ norm.*           &  1.0  & 1.1 & 1.5 & 1.9 & -   &  -  & -
\\
$K\mu\nu$ norm.*          &  3.2  & 4.8 & 3.9 & 4.8 & -   &  -  & -
\\
$X\ell\nu$ br. fr.$^\dag$      &  3.5  & 0.8 & 1.3 & 1.2 & 0.3 & 0.2 & 0.3
\\
$K^*\ell\nu$ f. f.$^\dag$      &  1.1  & 0.7 & 0.1 & 1.7 & 0.1 & 0.1 & 0.4 \\       
$c\bar{c}$ frag.$^\dag$   &  1.7  & 0.4 & 0.5 & 0.0   & 0.0  & 0.0  & 0.0   \\
$e$ mis-ID$^\dag$       &  0.4  & 0.2 & 0.5 & 0.4 & 0.0  & 0.0   & 0.0
\\
$\mu$ mis-ID$^\dag$        &  3.0  & 0.9 & 0.6 & 0.1 & 0.0  & 0.0  & 0.0
\\
\bbar\ norm.$^\dag$              &  0.2  & 0.2 & 0.1 & 0.2 & 0.0  & 0.0  & 0.0
\\\hline

Total (e)             &  6.7  & 4.9 & 5.8 & 7.5 & 0.7 & 0.7 & 2.7\\
Total ($\mu$)         &  8.0  & 6.9 & 6.9 & 8.6 & 0.7 & 0.7 & 2.7
\\\hline
Stat. ($e$)*           &  7.7  & 8.0 &10.3 &15.8 & 1.3 & 1.3 & 3.1 \\
Stat. ($\mu$)*       & 17.0  &24.6 &21.9 &28.7 & 4.8 & 3.7 & 4.3
\\\hline\hline

\end{tabular}
\raggedright \\
{*}Assumed uncorrelated across \qs\ bins.\\
\dag  Assumed correlated across \qs\ bins.\\
\ddag Assumed correlated across \qs\ bins in calculating $R_0$ and uncorrelated across \qs\ bins in the form factor fits.\\
\label{tab:syst}
\end{table}

Combining $R_{0_e}$ and $R_{0_\mu}$ after applying a $+1$\% correction to $R_{0_\mu}$ to account for the reduced muon phase space, gives
$$
R_0 = 0.082 \pm 0.006 \pm 0.005.
$$
This result is consistent with the previous world average of $0.101\pm 0.018$~\cite{PDG}, but is more precise.

We next determine parameters describing the form factors by fitting the corrected \qs\ distributions.  We first use a simple pole parameterization,
$$
f_+^{h}(q^2) = \frac{f_+^{h}(0)}{1-q^2/m_{\rm pole}^2},
$$
and vary the value of \mpole, constraining the integral over \qs\ to unity.  The quality of the fits is good.  Dominance by a single pole would imply $m_{\rm pole}^{D\to h} = m_{D_{(s)}^*}$.  We find $m_{\rm pole}^{D\to \pi}= 1.86^{+0.10+0.07}_{-0.06-0.03}$ \gev\ and $m_{\rm pole}^{D \to K} = 1.89\pm 0.05^{+0.04}_{-0.03}$  \gev, where the uncertainties are statistical and systematic. 
 We also fit the data with a modified pole distribution~\cite{BK}, 
$$
f_+^{h}(q^2) = \frac{f_+^{h}(0)}{(1-q^2/m^2_{D^*_{(s)}})(1-\alpha q^2/m^2_{D_{(s)}^*})},
$$
to obtain the parameter $\alpha$.  We find $\alpha^{D\to \pi} = 0.37^{+0.20}_{-0.31}\pm0.15$
 and $\alpha^{D\to K} = 0.36\pm 0.10^{+0.03}_{-0.07}$.  
Our results for $m_{\rm pole}^{D \to K}$ and $\alpha^{D\to K}$ suggest the existence of contributions beyond the pure $D_s^*$ pole to the \dtoklnu\ form factor.  For \dtopilnu,  $m_{\rm pole}^{D\to \pi}$ is consistent with the $D^*$ mass, though the precision is sufficient to rule out only large additional contributions. 

Several predictions for the form factors are superimposed on our data in Figure~\ref{fig:qsqrDist}.  Most are in satisfactory agreement with the data.  The ISGW2 model~\cite{isgw2}, however, predicts a \qs\ distribution for \dtoklnu\ that peaks lower than the data, and accordingly the $\chi^2$ with our data is poor (18 for 2 degrees of freedom). 

Using the value of \Rz\ and parameterizing the form factors with the results of the modified pole fit, we find 
$$
{|f^{\pi}_{+}(0)|^2 |V_{cd}|^2\over |f^K_{+}(0)|^2 |V_{cs}|^2} = 0.038^{+0.006+0.005}_{-0.007-0.003},
$$
where the uncertainties are statistical ($\pm 0.003$ from \Rz\ and $\pm0.006$ from $\alpha$) and systematic ($\pm 0.002$ from \Rz\ and $^{+0.004}_{-0.002}$ from $\alpha$).  The result is the same within 1\% if we use the simple pole form factor instead.   Using $|V_{cd}/V_{cs}|^2=0.052 \pm 0.001$~\cite{PDG}  gives $\rfo = 0.86\pm0.07^{+0.06}_{-0.04}\pm 0.01$, where the first error is statistical, the second is systematic, and the third is from the CKM matrix elements.  This value is consistent with most expectations for SU(3) symmetry breaking~[7-9,12].

We have presented a new measurement of the ratio of \dtopilnu\ to \dtoklnu\ decay rates.  This result is more precise than the previous world average by nearly a factor of two.   Our data also provide new information on the \dtoklnu\ form factor, a first determination of the \dtopilnu\ form factor and the first model independent constraint on $|f^{\pi}_{+}(0)| |V_{cd}|/ |f^K_{+}(0)| |V_{cs}|$.  Together, these offer new checks of SU(3) symmetry breaking and the form factors predicted for the semileptonic decays of heavy mesons into light ones.

We gratefully acknowledge the effort of the CESR staff 
in providing us with
excellent luminosity and running conditions.
This work was supported by 
the National Science Foundation,
the U.S. Department of Energy,
the Research Corporation,
and the 
Texas Advanced Research Program.

\end{document}